\documentclass[twocolumn,showpacs,amsmath,prl]{revtex4}
\usepackage{epsfig}
\usepackage{graphics}
\begin{document}

\title{Evidence for Landau's critical velocity in superfluid helium
nanodroplets from wave packet dynamics of attached potassium dimers}

\author{Martin Schlesinger$^1$}
 \email{Martin.Schlesinger@tu-dresden.de}
\author{Marcel Mudrich$^2$}
\author{Frank Stienkemeier$^2$}
\author{Walter T. Strunz$^1$}

\affiliation{$^1$Institut f\"ur Theoretische Physik, Technische Universit\"at Dresden, D-01062 Dresden, Germany}

\affiliation{$^2$Physikalisches Institut, Universit\"at Freiburg, D-79104 Freiburg, Germany}

\date{\today}
\begin{abstract}
Femtosecond pump-probe spectroscopy has been used to study vibrational
dynamics of
potassium dimers attached to superfluid helium nanodroplets.
Comparing the measured data with theoretical
results based on dissipative quantum dynamics we propose
that the most important effect of the helium environment is a general
damping of the vibrational dynamics as a result of the interaction
between dimer and collective degrees of freedom of the helium
droplet. The calculations allow us to explain crucial experimental
findings that are unobserved in gas-phase measurements. Remarkably, best
agreement with experiment is found for a model where we neglect damping
once a wave packet moves below a critical velocity. In
this way the results provide first direct evidence for the Landau
critical velocity in
superfluid nanodroplets.
\end{abstract}


\maketitle
Superfluidity in finite-sized quantum systems
is a fundamental issue of current interest, both in liquid helium (He)
\cite{Hartm_4560_1996, Grebe_2083_1998, Tang}
and in ultracold atomic gases \cite{Ketterle_}.
As first discussed by Landau, superfluidity manifests itself through the fascinating effect of
frictionless flow below a critical velocity~\cite{Landa_71_1941}.
In this Letter we present direct evidence for the existence of a critical
velocity in superfluid He nanodroplets.

Spectroscopy of probe molecules embedded inside He clusters
and He nanodroplets ($10^3-10^4$ atoms) has provided some insight into their
superfluid behaviour by studying
the response to external perturbations
\cite{Hartm_4560_1996, Grebe_2083_1998, Tang}.
In contrast to pure spectroscopic studies reported earlier, we present
femtosecond pump-probe measurements.
This technique is well-established for studying vibrational
wave packet (WP) dynamics
of diatomic systems \cite{Bowma_297_1989,Baume_639_1992,deViv_7789_1996,
Rutz_9_1997, Nicol_7857_1999}. Here, we investigate potassium molecules
($\text{K}_2$) attached to He
nanodroplets (He nanodroplet isolation (HENDI) spectroscopy)
\cite{Claas_1151_2006}.
He droplets provide a versatile test bed for studying relaxation dynamics
of the immersed species which are cooled to the droplet temperature
($0.38\text{ K}$) \cite{Toenn_2622_2004,Stien_R127_2006}.

The observed real time pump-probe spectra of K$_2$ on He droplets
differ significantly from previously obtained gas phase results
\cite{Claas_1151_2006, Nauta_9466_2000}.
In this paper we argue that dissipative
quantum dynamics, here employed through a quantum master
equation, is crucial for the understanding
of these measurements. Moreover, the dissipation has to be
combined with stochastic desorption of the molecule and
shifts of the potential energy surfaces induced by the He droplets
to correctly describe the experimental results.
Remarkably, we find best agreement with the experimental findings when
damping is omitted for very slowly moving WPs. This demonstrates the
potential of real time studies of vibrational
motions to investigate the Landau critical velocity on the microscopic scale,
very much in the spirit of vibrational wire resonators
\cite{Kraus_Erbe_Blick_2000} or quartz tuning forks \cite{Zadorozhko_2009}
that probe on larger scales.

\begin{figure}[ht]
\begin{center}
{
\includegraphics[width=0.49\textwidth]{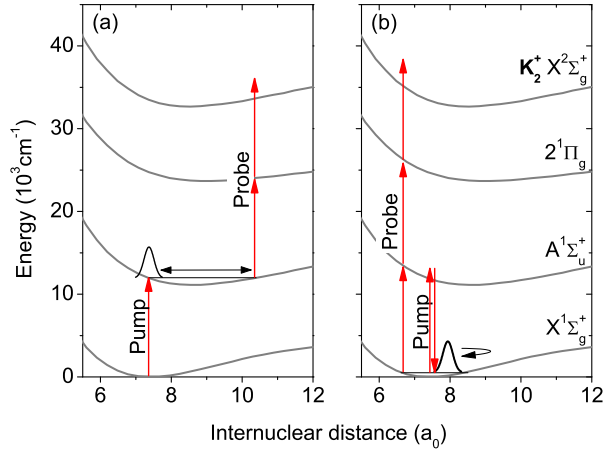}}
\caption{ \label{fig:Potentials}
Potential curves of the $\text{K}_2$ molecule and
excitation schemes at $\lambda=833$ nm (a) and at $\lambda=800$ nm (b).
The pump pulse creates vibrational WPs in various electronic states.
Significant ionization through the probe pulse occurs in (a) only 
if a WP is located around the outer,
and in (b) only if a wave packet is near the inner
turning point (FC window).}
\end{center}
\end{figure}

In the pump-probe spectroscopy of K$_2$, potassium atoms are attached to
He nanodroplets ($\sim 5\ 000$\,atoms) which are created in a
supersonic expansion of He gas at cryogenic conditions
\cite{Claas_1151_2006}.
Doping conditions are chosen such that on
average two potassium atoms stick to one droplet forming K$_2$ dimers which are
weakly bound on the surface of the droplets. In a one color pump-probe
excitation scheme the resulting K$_2^+$ photo ions are
recorded mass selectively. A laser having 110 fs pulse width has been
used providing 16 nJ pulses mildly focused in the interaction region
with the molecular beam. The (3)-photon excitation schemes which are
most relevant are depicted in Fig.~\ref{fig:Potentials}.

The time dependent Schr\"odinger equation for the state
$\Psi(t)=(\psi_{\rm X}(t),\psi_{\rm A}(t),\dots)$ is solved fully numerically,
as detailed for the gas phase in
Refs.~\cite{deViv_16829_1995,deViv_7789_1996}.
Here, $\psi_i(t)$ denotes the vibrational WP in electronic state $i$.
Transition dipole moments are assumed to be constant (Condon approximation).
In a first step we include
He-induced shifts of potential energy surfaces (PES).
Systematic numerical tests reveal that best agreement with experiment is
obtained when only shifting the $2\Pi$ surface (by $-50$cm$^{-1}$)
which agrees well with findings in
~\cite{Claas_1151_2006, Claas__2006}. As the excess energy of photo electrons
is not observed here, a shift of the ionic
surface as reported in \cite{Loginov_2005} has no effect on
our results.
Next, we take into account dimer desorption from the droplets.
We choose a constant desorption rate leading to
a probability density $P(t') \sim \exp\{-t'/\tau_D\}$ for the desorption time
$t'$ after electronic excitation.
Upon desorption, the He-induced shifts are set to zero.
We find that a description solely based on energy shift (parameter
$\Delta_{\Pi}$) and desorption (parameter $\tau_D$) cannot reproduce
crucial features of the HENDI measurements ~\cite{schle__2008}.

The new essential ingredient of our
description is the inclusion of damping of vibrational
WPs. Dissipation of vibrational energy
occurs due to the interaction with collective
degrees of freedom of the He droplets (phonons or ripplons).
In our phenomenological approach we do not specify the
microscopic dimer-droplet interaction. Instead,
we use a well-established master equation, describing damping
for near-harmonic systems at effectively zero temperature
fully quantum mechanically
(see \cite{Scull__1997}).
Thus, the dynamics of the full vibrational density operator
$\rho$ is determined from
\begin{equation}\label{eq:9}
\dot{\rho} = - \frac{i}{\hbar} [H,\rho]
+\frac{1}{2}\sum_i \gamma_i \left(2 a_i \rho a^{\dagger}_i
- a_i^\dagger a_i \rho - \rho a^\dagger_i a_i\right).
\end{equation}
Here, $H$ denotes the vibrational Hamiltonian
and $a_i,a^\dagger_i$ are the annihilation/creation
operators obtained from a harmonic approximation of the PES
corresponding to electronic state $i$.
We choose the parameters for damping (rates $\gamma_i\equiv\gamma $)
and desorption (time $\tau_D$) to be state-independent.
Note, however, a possible influence of the orientation of the dimer on
damping dynamics as discussed for Li$_2$ in \cite{Bovino_2009}. We
here assume no change of alignment during dynamics.

\begin{figure}[b]
\begin{center}
{
\includegraphics[width=0.49\textwidth,height=.305\textheight]{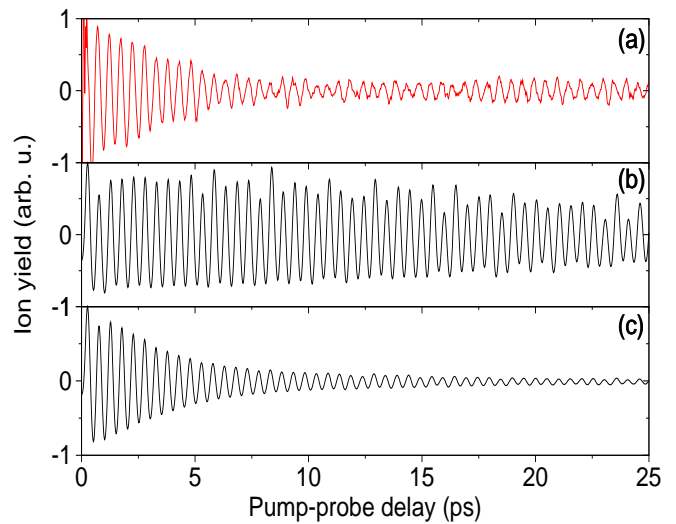}}
\caption{ \label{fig:PPTraces}
Pump-probe signal for $\lambda=833\text{ nm}$.
(a) Experimental HENDI result (from Ref.~\cite{Claas_1151_2006});
(b) Numerical simulation for the gas phase
(see also \cite{deViv_7789_1996});
(c) Calculated signal assuming He-induced damping, shift of PES and desorption of the molecule.
}
\end{center}
\end{figure}

Excitation at a wavelength of 833\ nm almost exclusively follows the path sketched in Fig.~\ref{fig:Potentials} (a) and probes the WP propagation in the $\text{A}^1\Sigma^+_u$ electronic state~\cite{Nicol_7857_1999}. The probe pulse leads to an enhanced population of the final ionic state only if the WP is
located at the outer turning point, where the intermediate $2\Pi$ state
opens a Franck-Condon (FC) window.
The resulting oscillatory structure
(Fig.~\ref{fig:PPTraces}) in the ion rate reflects the dynamics of the
WP in the A state. The simulation of the ion
signal for the gas phase is shown in Fig.~\ref{fig:PPTraces}(b) and agrees with
earlier gas phase results~\cite{deViv_7789_1996}. The weak overall decrease of
the oscillation amplitude in Fig.~\ref{fig:PPTraces}(b)
is caused by the spreading of the initially well-localized WP
in the anharmonic potential.
This gas phase result is markedly different from
the experimental HENDI spectrum (Fig.~\ref{fig:PPTraces} (a)),
which is a clear sign of the effect of the He environment.
The measured signal shows an exponential decay of the initial amplitude
which is well captured by our calculations (Fig. \ref{fig:PPTraces} (c)).
The persisting oscillations are also reproduced, yet with a smaller
amplitude.

Best agreement between experiment and our theory is obtained when
damping, shift and desorption are included into the calculation with
parameters $\gamma=0.15/\text{ps}$, $\Delta_{\Pi}=-50\text{ cm}^{-1}$
and $\tau_D=8\text{ ps}$ (Fig.~\ref{fig:PPTraces} (c)).
The dependence of all our findings
on these precise values is rather smooth and only after leaving a $\pm 5\%$
interval significant deviations from the displayed figures occur.

It is tempting to relate the damping rate $\gamma$
to measurements of the viscosity $\eta$ in liquid He (see
\cite{Zadorozhko_2009}). However, such values for $\eta$ are
based on macroscopic theories which are not applicable here.
Note that in \cite{Bovino_2009} the authors investigate
dissipative alkali dimer dynamics on a microscopic scale by
studying collisions with $^4$He atoms. Remarkably, the
calculated friction coefficient for
singlet systems ($0.06/$ps) turns out to be of the order of
magnitude of our $\gamma$.

\begin{figure}[b]
\begin{center}
{
\includegraphics[width=0.49\textwidth,height=0.17\textheight]{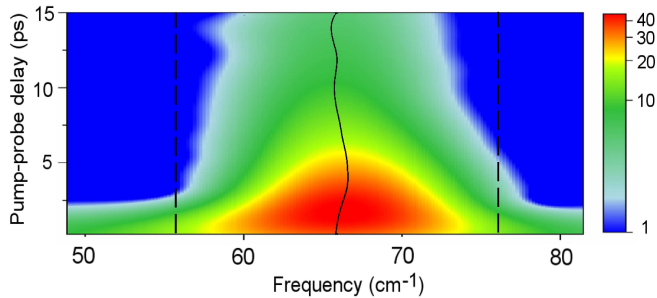}}
\caption{ \label{fig:shift}
Calculated pump-probe signal for $\lambda=833\text{ nm}$.
The dominant frequency is slightly shifted to larger values for
short delay times (the black solid line indicates the
the mean frequency between the dashed lines as a function of time).
}
\end{center}
\end{figure}

The temporal evolution and intensity of certain frequency components can be
visualized by employing sliding window Fourier transforms (FT) on the
time-dependent signal (``spectrogram'' \cite{Rutz_9_1997}).
As discussed in Ref.~\cite{Claas_1151_2006}, in the experiment at 833\,nm
excitation wave length a small increase of the frequency of the WP motion
during the first 10 ps is
observed. In Fig. \ref{fig:shift}
this behavior (if less pronounced) is seen in the FT spectrum of our
calculated signal, provided damping is included.
The findings allow the following
interpretation:
during the first few picoseconds the contribution of the damped
molecules is most important.
As a consequence of the anharmonic potential, they
vibrate with a higher frequency, explaining the
initial shift of the main frequency $\omega_A^{\scriptsize 833\text{nm}}$
to slightly larger values (or lower vibrational
quantum numbers). Later on, for times $\tau>10\text{ ps}$,
vibrationally damped molecules no longer contribute, as a closing of
their FC window takes place.
Hence, only molecules which desorb very early and thus do not suffer
vibrational damping contribute to the ion yield at late delay times.
Consequently, a weak oscillatory signal
persists in the time domain and in the windowed FT the main frequency
returns to the original gas phase value.

\begin{figure}[ht]
\begin{center}
{
\includegraphics[width=0.41\textwidth,height=0.55\textheight]{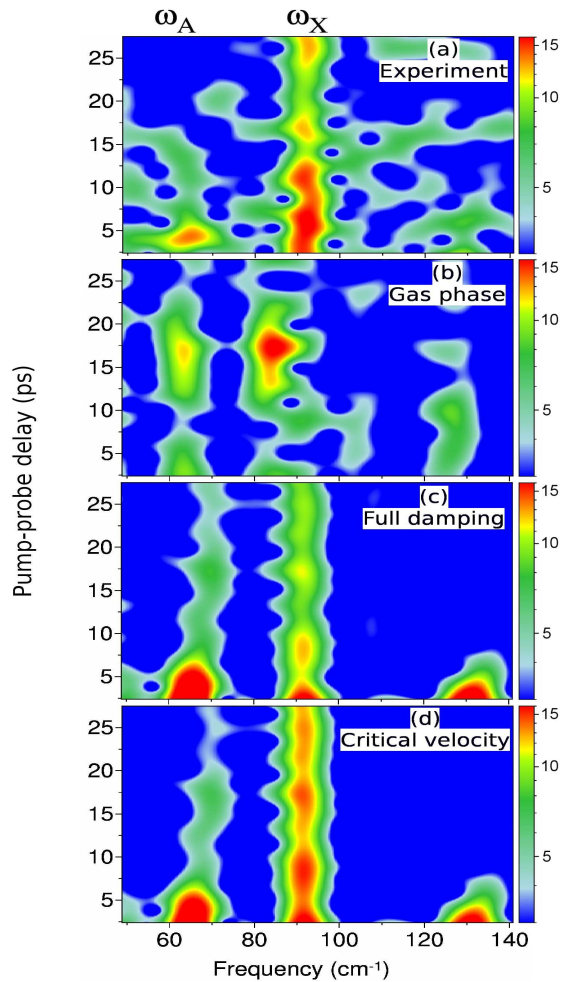}}
\caption{ \label{fig:spectro}
      Spectrograms showing the WP dynamics at $\lambda=800$ nm for the experimental HENDI result (a), for the gas-phase simulation (b), for the fully damped model (c), and for simulated data including a critical velocity (d).}
\end{center}
\end{figure}

At the shorter wave length $\lambda=800$ nm the excitation scheme
sketched in Fig.~\ref{fig:Potentials} (b) applies. Here, a WP in the
ground state, excited through
stimulated Raman scattering (RISRS), can be visualized through 3-photon
absorption. In the gas phase, this process is superposed coherently with
transitions from higher lying electronic states near their respective inner
turning points,
leading to interferences in the ionic state.
A more detailed analysis shows that in the gas phase the interference is indeed
destructive: the contribution of the WP on the X
surface is missing
in the gas phase (see Fig.~\ref{fig:spectro} (b))
-- see similar findings in
\cite{baumert1992}. Without any damping mechanism,
contributions from the A state
(component $\omega_A^{\scriptsize 800\text{nm}}\approx 64$cm$^{-1}$)
and from the $2 \Pi$ state
(component $2\,\omega_{2\Pi}^{\scriptsize 800\text{nm}}\approx 86$cm$^{-1}$)
are dominant,
which is not observed in the He droplet experiment (Fig.~\ref{fig:spectro}(a)). There, although
$\omega_A^{\scriptsize 800\text{nm}}$ is present during the first 5 ps,
it fades out and
only the component $\omega_X^{\scriptsize 800\text{nm}}\approx 91$ cm$^{-1}$
(WP in X state) contributes at later delay times.
We now include damping for
$\lambda=800\text{ nm}$
with precisely the same
parameters as for the considerations
at $\lambda=833\text{ nm}$.
The corresponding spectrogram (Fig.~\ref{fig:spectro}(c)) agrees remarkably
well with experimental findings in many respects.
The component at $\omega_A^{\scriptsize 800\text{nm}}$
fades out after several ps
due to vibrational damping: the WP in the A state leaves the
initial FC window at the inner turning point.

More significant
is the behavior of the  component near $\omega_X^{\scriptsize 800\text{nm}}$,
ascribed to the ground state WP. It is
clearly visible in the experimental and theoretical (including damping)
spectrogram, yet absent in the gas phase.
Damping of the WPs in the A and
$2\Pi$ state enhances ionization from the ground state by the probe
pulse, since the mapping
of the X state dynamics to the ion no longer suffers
from destructive interference with
competing processes near the inner turning point on the A and $2\Pi$ surfaces.
\begin{figure}[t]
\begin{center}
{
\includegraphics[width=0.49\textwidth]{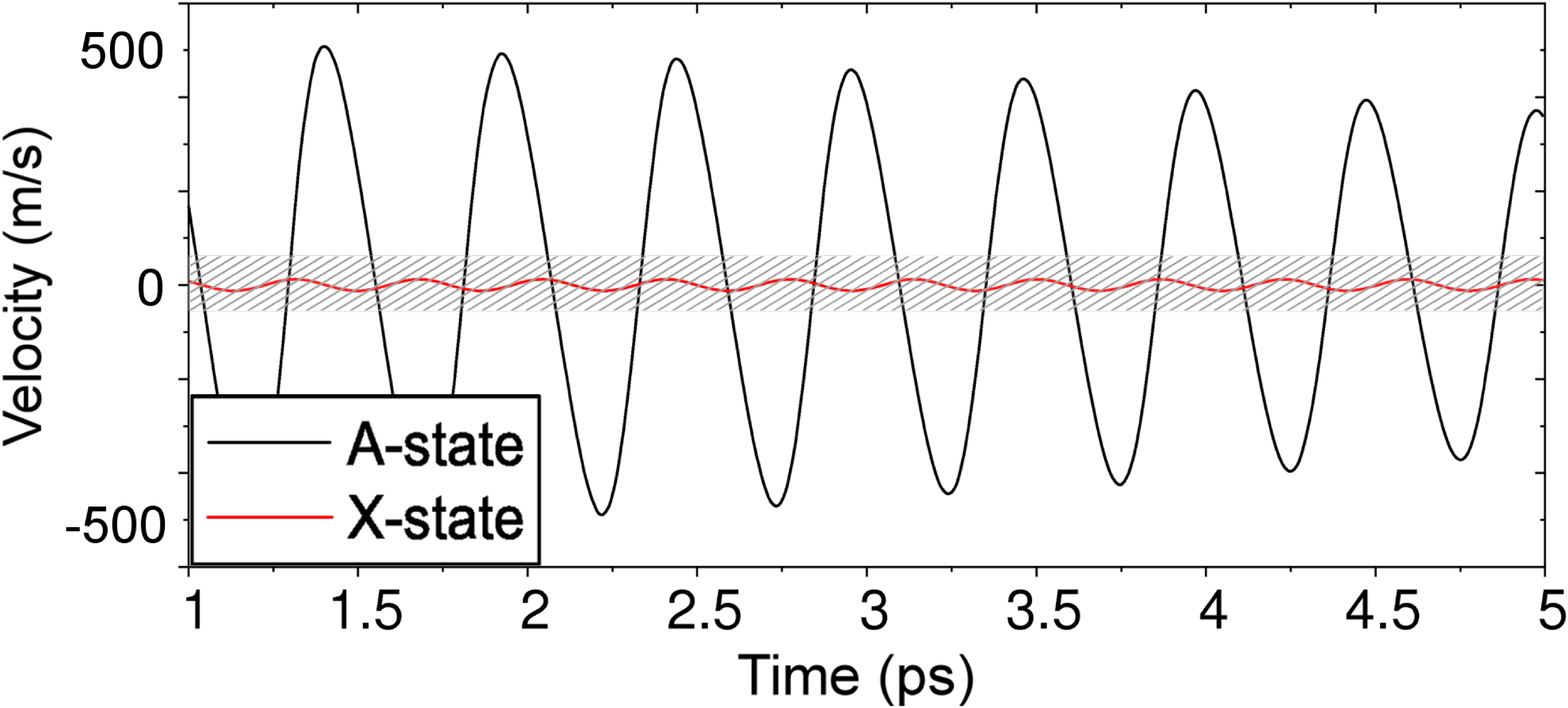}}
\caption{ \label{fig:velo}
Dynamics in velocity domain of a WP excited in the electronically excited $A$-state and in the ground state $X$. The hatched area indicates velocities below 
the critical value of 60 m/s.
}
\end{center}
\end{figure}
Most remarkably, the results at 800 nm hint at a direct influence of superfluidity on these spectra.
Molecules are assumed to be able to move unhindered through a superfluid
as long as their velocity does not exceed the (Landau) critical
velocity $v_c$ \cite{Landa_71_1941}. In a simple approach we determine
from our calculations the average velocity $\langle v
\rangle \ = \langle \hat{P} \rangle /2 \mu$ of the nuclear WP.
Once it drops below the critical velocity
(we chose the bulk value $v_c=60\text{ m/s}$),
we set the friction coefficient to zero
(see also \cite{Bonho_11359_2004}) and compare with the fully damped model.

Fig.~\ref{fig:velo} illustrates the WP velocities in the ground
($X$) and excited state ($A$). For the excitation at
$\lambda = 833$ nm, it is the dynamics on the A surface that
dominates the observations. This vibration, however, is fast and the
corresponding
velocity oscillates between $\pm 500$ m/s. For most of the time, therefore,
friction is
present and the influence of $v_c$ is hardly noticeable. For an
excitation at wave length $\lambda = 800$ nm, however,
the motion on the X surface is so slow (at most $v\approx 15$ m/s) that we
neglect damping on that surface entirely ($\gamma_X=0$).
In the resulting
spectrogram (Fig.~\ref{fig:spectro}(d)), the component
$\omega_X^{\scriptsize 800\text{nm}}$ is even more pronounced
compared to the previous calculation (full damping). Most strikingly, the
agreement with the experimental result is improved.

In conclusion, the study of vibrational wave packets in dimers attached
to He
nanodroplets exhibits clear deviations from previous gas phase results.
A model based
on shifts of potential energy surfaces, desorption from the host and,
most importantly, damping
of the wave packets is able to
explain crucial features of the measurement. For excitation
at $\lambda=833\text{ nm}$ the decay of the
signal and a small shift in the FT spectrum are reproduced.
Damping of excited state wave packets enhances
the mapping of the ground state dynamics to the ion signal, as realized at
$\lambda=800\text{ nm}$.

Most remarkably, in our studies best agreement with experiment is obtained
when we neglect damping for very slowly moving wave packets.
We attribute this observation to the role of the critical velocity in
these experiments, which turns out to be larger than the velocity of
the motion on the X surface. Thus, the influence
of superfluidity is directly
visible, opening the door for further studies:
choosing heavier molecules for which the vibrational
velocities are well below some 100\,m/s, in combination with a microscopic
fully quantum calculation will enable us to investigate
friction and frictionless motion on the atomic scale.

We thank Jan Handt and Ralf Sch\"utzhold for stimulating discussions
and C. P. Schulz for experimental aid.
Support by the Deutsche Forschungsgemeinschaft (DFG) is
gratefully acknowledged.
Computing resources have been provided by the
{\it Zentrum f\"ur Informationsdienste und Hochleistungsrechnen} (ZIH) at
the TU Dresden. M. Schlesinger is a member of the IMPRS Dresden.
\bibliographystyle{apsrev}

\end{document}